\documentclass[aps,pre,reprint,amsmath,amssymb,superscriptaddress,nofootinbib]{revtex4-2}

\usepackage{bm}
\usepackage{graphicx}
\usepackage{microtype}
\usepackage[colorlinks=true,linkcolor=blue,citecolor=blue,urlcolor=blue]{hyperref}

\newcommand{\sech}{\operatorname{sech}}
\newcommand{\ii}{\mathrm{i}}
\newcommand{\dd}{\mathrm{d}}
\newcommand{\cH}{\mathcal{H}}

\newcommand{\cJ}{\mathcal{J}}

\begin{document}

\title{Divergence Without Transition in Adiabatic Theory: Exact Cancellation in Reflectionless Potentials}

\author{Amitava Bhattacharjee}
\affiliation{Department of Astrophysical Sciences, Princeton University, Princeton, New Jersey 08544, USA}

\date{July 11 2026}

\begin{abstract}

Adiabatic invariants play a central role in plasma physics, from
magnetic moment and bounce action to wave action in slowly varying
media. Their perturbative constructions are often asymptotic, and
exhibit factorial growth. We show that such divergence does not by
itself imply non-adiabatic transitions. For the reflectionless
potential hierarchy associated with Korteweg--de Vries solitons,
the exact backward-wave coefficient vanishes, although Berry's
phase-integral iteration and the corresponding Lie-transform
construction are divergent. Darboux factorisation gives the
transmitted wave explicitly. Its modulus and phase define a
normal form in which the moving canonical frame is distorted inside
the interaction region but returns to its original asymptotic form,
leaving only a phase shift and no action change. The exact phase
integral is nevertheless an unstable fixed point of the derivative
iteration. For the one-soliton symmetry point, an explicit Borel calculation exhibits nonzero 
singularities in an individual Lie-transform family even though the exact off-diagonal connection 
coefficient vanishes. Analyticity of the exact connection data then requires these representation-dependent ambiguities to cancel in the completed connection matrix. Thus
divergence diagnoses failure of local diagonalisation, whereas the
global symplectic connection determines whether reflection survives.

\end{abstract}

\maketitle

\section{Introduction}

Adiabatic invariance entered modern plasma physics through nearly periodic particle motion. The magnetic moment of a rapidly gyrating particle, the longitudinal bounce action of a mirror-trapped particle, and the action of a slowly modulated plasma wave are conserved to successive algebraic orders when equilibrium fields vary on scales long compared with the fast orbit or wave period. Kruskal placed these examples in a general Hamiltonian framework, proving the existence of formal adiabatic invariants to all orders for systems whose trajectories are nearly periodic \cite{Kruskal1962}. Guiding-center and bounce-center theory remain direct descendants of this construction \cite{Northrop1963,CaryBrizard2009}.

The phrase ``to all orders,'' however, does not imply convergence. The coefficients generated by repeated near-identity transformations usually grow factorially, because each additional order differentiates the slowly varying coefficients and probes their nearest singularities in the complex plane. This is a characteristic example of what Kruskal called \emph{asymptotology}: the exact dynamics may be regular even when the natural formal construction is divergent \cite{Kruskal1963}. The physical interpretation of that divergence can be subtle. In generic analytic problems, optimal truncation leaves an exponentially small remainder, and that remainder often coincides with weak mode conversion, wave reflection, or violation of an adiabatic invariant.

Berry's iterative scheme for calculating the geometric phase to higher orders made this connection especially vivid \cite{Berry1987,Berry1990Histories}. Successive changes of basis reduce the local coupling between positive- and negative-frequency branches until an optimal order is reached, after which the sequence diverges and an exponentially small transition becomes visible. The classical counterpart of this program was initiated by Bhattacharjee and Sen, who constructed the Lie-transform expansion
of finite-rate corrections to the Hannay angle for the generalized harmonic oscillator---precisely the class of systems treated here---and noted the interpretational ambiguity of separating such angles into dynamical and geometric parts at finite rate \cite{SenBhatt1988}. Gjaja and Bhattacharjee (GB) proved that the Lie-transform expansion, like Berry's iteration, diverges factorially \cite{GjajaBhattacharjee1990}. Both constructions exhibit $k!\epsilon^k$ growth. GB concluded that divergence belongs to the asymptotic procedure and does not, by itself, prove a physical transition or a failure of action conservation.

The strongest test of this distinction is a system that is exactly reflectionless. Such systems arise naturally in the soliton theory that grew out of plasma physics. Zabusky and Kruskal coined the word \emph{soliton} while studying the Korteweg--de Vries (KdV) equation as a model of nonlinear waves in a collisionless plasma \cite{ZabuskyKruskal1965}. The inverse-scattering solution of KdV then identified reflectionless Schr\"odinger operators as the spectral data of pure solitons \cite{GardnerEtAl1967}. 

GB constructed analytic reflectionless oscillator histories for which the original iterative and perturbative schemes nevertheless neither terminate nor converge absolutely \cite{GjajaBhattacharjee1992}. In related work, Berry and Howls explained the local complex-plane mechanism of reflectionlessness \cite{BerryHowls1990}. The relevant turning points do not act independently: they form a tight zero--pole cluster whose exact comparison equation has half-integer Bessel solutions, or ``fake Airy functions,'' with terminating asymptotics and no ordinary Airy-type Stokes switching. At first sight this seems to conflict with the divergent adiabatic constructions obtained by GB.

The purpose of this paper is to reconcile these different perspectives. The common reduced equation is the slowly varying oscillator or one-dimensional wave equation. The reflected spatial wave, the negative-frequency temporal branch, and the change of a classical action are different manifestations of the same phenomenon. We show that three representations of the reflectionless problem — the finite Darboux dressing, the exactly closing Ermakov frame, and the divergent local adiabatic expansion — have sharply different asymptotic behavior.

Our main result is exact. For the generalized P\"oschl--Teller problem the backward-wave coefficient is proportional to $\sin(\pi\lambda)$,
where $\lambda(\lambda+1)$ measures the potential strength, and therefore vanishes at every integer $\lambda$. A Darboux representation then produces the transmitted solution, while its modulus and phase supply an Ermakov structure that closes asymptotically, leaving only a phase holonomy. Nevertheless, in Berry's iteration scheme, the exact phase-integral function is an unstable fixed point because the linearized map contains an unbounded differential operator. Thus the same system manifests a divergent all-orders construction and exactly zero mode conversion.

A closely related nonperturbative perspective has recently been developed by Updike and Burby for charged-particle motion in a
homogeneous, time-dependent magnetic field \cite{UpdikeBurby2026}. Building on the exact magnetic-moment invariant of Qin and Davidson
\cite{QinDavidson2006}, they show that this invariant can be constructed to be asymptotic to Kruskal's formal adiabatic series,
the required branch of the underlying Ermakov--Pinney solution being selected by a Bremmer series that organizes the associated
Hill-equation solution as a convergent sum over successive reflections. In contrast to the formal series, the exact invariant
retains information about parametric resonances, which decide whether conservation actually fails over long times. Their result
and the present one address complementary aspects of the same question: an exact invariant may exist despite formal divergence,
while actual growth or mode conversion is decided only by the global evolution map---by the Floquet multiplier in their periodic setting,
by the Stokes connection coefficient here.

\section{Formulation}

Consider
\begin{equation}
\epsilon^2\psi''(x)+P^2(x,\epsilon)\psi(x)=0,
\label{eq:wave-equation}
\end{equation}
where $P^2$ is real and positive on the real axis and approaches unity as $x\rightarrow\pm\infty$. Equation~\eqref{eq:wave-equation} is a universal local normal form.

If $x$ is time, it describes a harmonic oscillator with a slowly varying frequency. This is the elementary model for gyro, bounce, or trapping motion when the restoring force is linearized about a stable orbit. If $x$ is space, it describes above-barrier propagation of a scalar wave through a slowly varying medium. The two WKB branches represent forward and backward waves, or equivalently positive- and negative-frequency components. In a coupled-mode formulation, their mixing is mode conversion.

For a classical oscillator, the instantaneous action is
\begin{equation}
\cJ_{\rm inst}=\frac{1}{2P}\left[p^2+P^2q^2\right],
\label{eq:instant-action}
\end{equation}
with $p=\epsilon q'$. In magnetized-particle language this is the prototype of the magnetic moment or bounce action. For a wave it is the prototype of wave action. Reflectionlessness is therefore not merely an optical property: it is an exactly transitionless adiabatic cycle.

\section{Exact similarity and the adiabatic invariant}

Let a complex solution be written as
\begin{equation}
\psi(x)=\rho(x)\exp\left[\frac{\ii}{\epsilon}S(x)\right],
\qquad \rho(x)>0.
\label{eq:polar-solution}
\end{equation}
Choosing the current normalization $\rho^2S'=1$, the real and imaginary parts of Eq.~\eqref{eq:wave-equation} give
\begin{equation}
S'(x)=\frac{1}{\rho^2(x)},
\label{eq:phase-rate}
\end{equation}
\begin{equation}
\epsilon^2\rho''+P^2\rho=\frac{1}{\rho^3}.
\label{eq:ermakov}
\end{equation}
Equation~\eqref{eq:ermakov} is the Ermakov--Milne--Pinney equation \cite{Ermakov1880,Pinney1950}. Lewis obtained the corresponding exact invariant by applying Kruskal's all-orders asymptotic theory to the oscillator and recognizing that, in this special case, the formal construction can be summed in closed form \cite{Lewis1968}.

Define
\begin{equation}
\psi(x)=\rho(x)Q(T),
\qquad
\frac{\dd T}{\dd x}=\frac{1}{\rho^2(x)}.
\label{eq:similarity-map}
\end{equation}
Then Eq.~\eqref{eq:wave-equation} reduces exactly to
\begin{equation}
\epsilon^2\frac{\dd^2Q}{\dd T^2}+Q=0.
\label{eq:unit-oscillator}
\end{equation}
The original dynamics is a rigid rotation in a moving phase-space frame. The exact Lewis--Ermakov invariant is
\begin{equation}
\cJ_{\rm E}=\frac{1}{2}\left[\left(\frac{q}{\rho}\right)^2+
\left(\rho p-\epsilon\rho' q\right)^2\right].
\label{eq:Lewis-invariant}
\end{equation}
Unlike $\cJ_{\rm inst}$, Eq.~\eqref{eq:Lewis-invariant} is conserved exactly.

In canonical variables $\bm z=(q,p)^T$, the evolution factorizes as
\begin{equation}
\bm z(x)=\mathsf S_\rho(x)\,
\mathsf R\left(\frac{T(x)-T(x_0)}{\epsilon}\right)
\mathsf S_\rho^{-1}(x_0)\bm z(x_0),
\label{eq:symplectic-factorization}
\end{equation}
where
\begin{equation}
\mathsf S_\rho(x)=
\begin{pmatrix}
\rho & 0\\
\epsilon\rho' & \rho^{-1}
\end{pmatrix}.
\label{eq:squeeze-matrix}
\end{equation}
The matrix $\mathsf S_\rho$ dilates and shears phase space while preserving area. Reflectionlessness is 
the global condition that the asymptotic frames match:
\begin{equation}
\rho(+\infty)=\rho(-\infty),
\qquad
\rho'(+\infty)=\rho'(-\infty)=0.
\label{eq:clean-asymptotics}
\end{equation}

\section{Darboux representation}

The oscillator histories we analyze are the family
$P^{2}(x,\epsilon)=1+\epsilon^{2}m(m+1)\,\mathrm{sech}^{2}x$, whose
$m=1$ member is precisely the reflectionless example
$\omega^{2}=1+2\epsilon^{2}\,\mathrm{sech}^{2}\tau$ constructed in
Ref.~[11]. Writing $k=\epsilon^{-1}$, Eq.~(1) for this family is the
stationary scattering problem at energy $k^{2}$ for the
P\"oschl--Teller operators $H_{m}$ below---the reflectionless
Schr\"odinger operators that inverse-scattering theory identifies as
the spectral data of pure $m$-soliton solutions of KdV [10]. Their
distinguishing algebraic property is that they form a ladder:
$H_{m}$ is connected to $H_{m-1}$, and hence ultimately to the free
operator $H_{0}$, by first-order Darboux representations. This is what
makes the connection problem solvable \emph{finitely}: the one-way
solution of $H_{m}$ can be obtained from the free wave
$e^{ikx}$ by $m$ applications of a first-order differential
operator. The construction is as follows.

Introduce the Schr\"odinger operators
\begin{equation}
\cH_m=-\frac{\dd^2}{\dd x^2}-m(m+1)\sech^2x,
\qquad m=0,1,2,\ldots,
\label{eq:Hm}
\end{equation}
and the first-order operators
\begin{equation}
\mathsf A_m^\dagger=-\frac{\dd}{\dd x}+m\tanh x,
\label{eq:Darboux-operator}
\end{equation}
which satisfy the relation
\begin{equation}
\cH_m\mathsf A_m^\dagger=\mathsf A_m^\dagger\cH_{m-1}.
\label{eq:intertwining}
\end{equation}
Starting from a free right-going wave, one obtains the exact solution
\begin{equation}
f_m^+(x,k)=
\frac{
\displaystyle\prod_{n=1}^{m}
\left(n\tanh x-\frac{\dd}{\dd x}\right)e^{\ii kx}
}{
\displaystyle\prod_{n=1}^{m}(n-\ii k)
},
\qquad k=\epsilon^{-1}.
\label{eq:Darboux-Jost}
\end{equation}
The product is finite. Moreover,
\begin{equation}
f_m^+(x,k)\sim e^{\ii kx},
\qquad x\rightarrow+\infty,
\label{eq:right-Jost-plus}
\end{equation}
while
\begin{equation}
f_m^+(x,k)\sim
(-1)^m\prod_{n=1}^{m}\frac{n+\ii k}{n-\ii k}
\,e^{\ii kx},
\qquad x\rightarrow-\infty.
\label{eq:right-Jost-minus}
\end{equation}
No $e^{-\ii kx}$ branch is generated. The left-incident transmission amplitude is
\begin{equation}
t_m(k)=\prod_{n=1}^{m}\frac{k+\ii n}{k-\ii n},
\qquad
r_m(k)=0,
\label{eq:Darboux-scattering}
\end{equation}
with phase
\begin{equation}
\delta_m(k)=2\sum_{n=1}^{m}\arctan\left(\frac{n}{k}\right)
\quad (\mathrm{mod}\;2\pi).
\label{eq:transmission-phase}
\end{equation}
Equations~\eqref{eq:Darboux-Jost}--\eqref{eq:Darboux-scattering} are the finite Darboux representation of the pure soliton hierarchy \cite{Darboux1882,Crum1955,KayMoses1956}.

\section{Exact global connection coefficient}

To expose the cancellation away from integer coupling, consider
\begin{equation}
\psi''+\left[k^2+\lambda(\lambda+1)\sech^2x\right]\psi=0,
\label{eq:general-PT}
\end{equation}
with arbitrary real $\lambda$. Let
\begin{equation}
z=\frac{1-\tanh x}{2}.
\label{eq:z-variable}
\end{equation}
A right solution is
\begin{equation}
f_\lambda^+(x,k)=e^{\ii kx}
{}_2F_1\left(-\lambda,\lambda+1;1-\ii k;z\right),
\label{eq:hypergeometric-Jost}
\end{equation}
which tends to $e^{\ii kx}$ as $x\rightarrow+\infty$. The hypergeometric connection formula at $z=1$ gives (Appendix A)
\begin{equation}
f_\lambda^+(x,k)\sim
A_\lambda(k)e^{\ii kx}+B_\lambda(k)e^{-\ii kx},
\qquad x\rightarrow-\infty,
\label{eq:Jost-connection}
\end{equation}
where
\begin{equation}
A_\lambda(k)=
\frac{\Gamma(1-\ii k)\Gamma(-\ii k)}
{\Gamma(1+\lambda-\ii k)\Gamma(-\lambda-\ii k)},
\label{eq:A-lambda}
\end{equation}
\begin{equation}
B_\lambda(k)=
\frac{\Gamma(1-\ii k)\Gamma(\ii k)}
{\Gamma(-\lambda)\Gamma(1+\lambda)}.
\label{eq:B-gamma}
\end{equation}
For a unit-amplitude wave incident from the left, $t_\lambda=1/A_\lambda$ and $r_\lambda=B_\lambda/A_\lambda$; the zero of $B_\lambda$ is therefore exactly the zero of the physical reflection amplitude.

\begin{figure*}[t]
\includegraphics[width=0.91\textwidth]{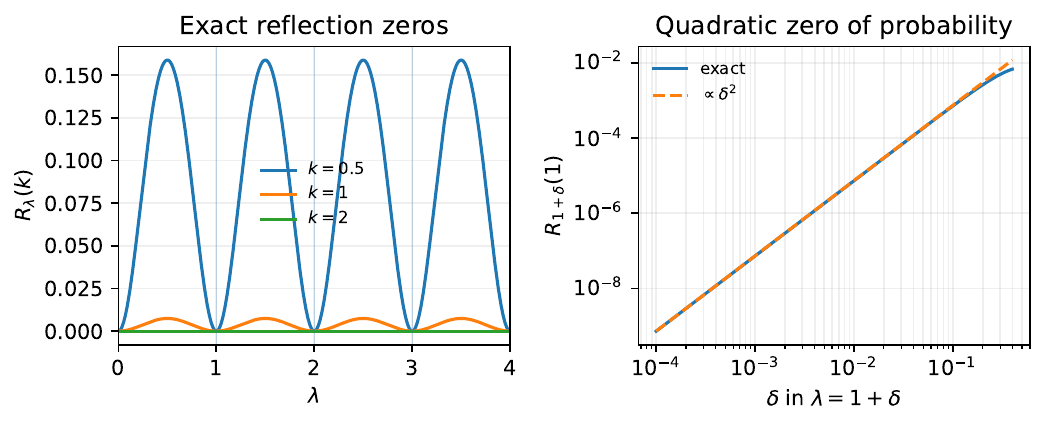}
\caption{Exact cancellation of the backward-wave or mode-conversion coefficient. Left: the reflection probability in Eq.~\eqref{eq:R-exact} vanishes at every integer $\lambda$, for all $k$. Right: near $\lambda=1$, the probability has a quadratic zero, Eq.~\eqref{eq:R-detuning}; the underlying amplitude has a simple zero.}
\label{fig:reflection-zeros}
\end{figure*}

Using the gamma-function reflection identity yields the exact result (Appendix A)
\begin{equation}
B_\lambda(k)=\ii\frac{\sin(\pi\lambda)}{\sinh(\pi k)}.
\label{eq:B-exact}
\end{equation}
The physical reflection probability is therefore
\begin{equation}
R_\lambda(k)=
\frac{\sin^2(\pi\lambda)}
{\sinh^2(\pi k)+\sin^2(\pi\lambda)}.
\label{eq:R-exact}
\end{equation}
At integer coupling,
\begin{equation}
\lambda=m\in\mathbb N,
\qquad
B_m(k)=R_m(k)=0
\label{eq:integer-zero}
\end{equation}
for every real $k$. Near an integer, $\lambda=m+\delta$,
\begin{equation}
B_{m+\delta}(k)=
\ii(-1)^m\frac{\pi\delta}{\sinh(\pi k)}+O(\delta^3),
\label{eq:B-detuning}
\end{equation}
\begin{equation}
R_{m+\delta}(k)=
\frac{\pi^2\delta^2}{\sinh^2(\pi k)}+O(\delta^4).
\label{eq:R-detuning}
\end{equation}
Figure~\ref{fig:reflection-zeros} displays these exact zeros and the quadratic detuning law.

For $k\gg1$,
\begin{equation}
B_\lambda(k)\sim
2\ii\sin(\pi\lambda)e^{-\pi k}.
\label{eq:B-asymptotic}
\end{equation}
The singulant $\pi k=\pi/\epsilon$ remains present, but the multiplier of the opposite branch vanishes exactly at integer $\lambda$. In WKB language, Eq.~\eqref{eq:B-exact} provides the global off-diagonal connection. Its zero is the precise observable statement of Stokes cancellation. It is not necessary, and generally not basis invariant, to assign zero separately to every local matrix used in a decomposition of the cluster. A coefficient-level realization of this same exponential scale,
for the one-soliton Lie-transform expansion, is given in
Appendix~\ref{app:gb-borel}.

\section{The one-soliton solution}

For $m=1$, a left-incident unit-amplitude solution is
\begin{equation}
\psi_{\rm in}(x,k)=
e^{\ii kx}\frac{k+\ii\tanh x}{k-\ii}.
\label{eq:m1-Jost}
\end{equation}
Writing $k=\epsilon^{-1}$ and using Eq.~\eqref{eq:polar-solution}, one obtains
\begin{equation}
\rho^2(x)=
\frac{1+\epsilon^2\tanh^2x}{1+\epsilon^2}
\label{eq:m1-rho}
\end{equation}
and
\begin{equation}
S'(x)=\frac{1}{\rho^2(x)}
=\frac{1+\epsilon^2}{1+\epsilon^2\tanh^2x}.
\label{eq:m1-phase-rate}
\end{equation}

\begin{figure*}[t]
\includegraphics[width=0.91\textwidth]{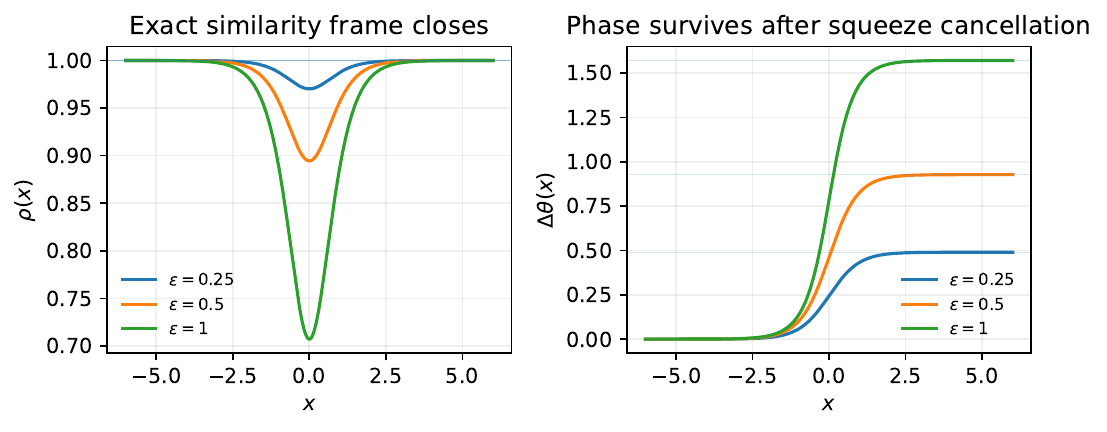}
\caption{Exact Ermakov closure for the one-soliton. Left: the scale factor in Eq.~\eqref{eq:m1-rho} is deformed within the interaction region but returns to $\rho=1$ with zero derivative. Right: the additional phase $\Delta\theta=(S-x)/\epsilon$ approaches the nonzero transmission phase $2\arctan\epsilon$. Reflectionlessness removes the net squeeze, not the phase holonomy.}
\label{fig:similarity-closure}
\end{figure*}

A convenient phase convention is
\begin{equation}
\frac{S(x)-x}{\epsilon}
=
\arctan(\epsilon\tanh x)+\arctan\epsilon .
\end{equation}
Consequently,
\begin{equation}
\rho(\pm\infty)=1,
\qquad
\rho'(\pm\infty)=0,
\label{eq:m1-closure}
\end{equation}
while the surviving phase is
\begin{equation}
\delta_1=2\arctan\epsilon.
\label{eq:m1-delta}
\end{equation}
Figure~\ref{fig:similarity-closure} shows the distinction directly: the similarity scale returns to unity, whereas the phase retains a finite holonomy. The character of this surviving phase deserves emphasis. The pulse $P^{2}(x,\epsilon)$ rises from unity and returns along the same path, so the parameters traverse a retraced loop enclosing zero area, and the adiabatic Hannay angle vanishes identically. The
holonomy $\delta_{m}=2\sum_{n=1}^{m}\arctan(n\epsilon) = m(m+1)\,\epsilon+O(\epsilon^{3})$ is therefore a purely finite-rate angle: it is exactly the object whose Lie-transform expansion was constructed in Ref.~\cite{SenBhatt1988}. For the reflectionless hierarchy, that expansion possesses an elementary closed sum, analytic in $\epsilon$ with radius of convergence $1/m$ set not by the turning-point cluster of Sec.~VII but by the
deepest soliton bound state at $k=im$, i.e., by the poles of the transmission amplitude~(18).

In the temporal-oscillator interpretation, the asymptotic positive- and negative-frequency amplitudes obey a Bogoliubov transformation. In the quantum oscillator this is the transformation of creation and annihilation amplitudes:
\begin{equation}
a_{\rm out}=\alpha a_{\rm in}+\beta a_{\rm in}^\dagger.
\label{eq:Bogoliubov}
\end{equation}
The off-diagonal coefficient is the reflected-wave coefficient, and Eqs.~\eqref{eq:integer-zero} and \eqref{eq:m1-closure} imply
\begin{equation}
\beta=0,
\qquad
\alpha=e^{\ii\delta_m}.
\label{eq:no-particle-creation}
\end{equation}
For the one-soliton case, Appendix~\ref{app:gb-borel}
shows how this cancellation appears inside the original
GB Lie-transform expansion. The state can be strongly squeezed relative to the instantaneous basis
at intermediate $x$, but the squeeze is undone asymptotically.  The
reflectionless evolution is therefore cyclic in the noncompact part of
the canonical transformation: the endpoint frames agree, while the
compact angle variable advances by the transmission phase
\begin{equation}
\delta_m
=
2\sum_{n=1}^{m}\arctan(n\epsilon).
\end{equation}
Thus the finite-rate angle computed in the Hannay-angle formulation is
not a residual squeeze or action change, but the surviving phase
holonomy.

For a scattering history, the corresponding global object is the
asymptotic connection matrix
\begin{equation}
\mathsf{C}
=
\begin{pmatrix}
\alpha & \beta^{*}\\
\beta & \alpha^{*}
\end{pmatrix},
\qquad
|\alpha|^{2}-|\beta|^{2}=1 .
\end{equation}
Here $\beta$ measures reflection or positive--negative-frequency
conversion.  In the reflectionless Darboux hierarchy,
\begin{equation}
\beta=0,
\end{equation}
although the diagonal phase of $\alpha$ remains nontrivial.  Thus the
periodic and scattering problems are unified by the same symplectic
normal-form geometry, but are distinguished by the global holonomy.

\section{The Berry--Howls cluster}

The nearest poles of $\sech^2x$ occur at
\begin{equation}
x_*=\frac{\ii\pi}{2}+\ii\pi n.
\label{eq:complex-poles}
\end{equation}
Near $x_*=\ii\pi/2$,
\begin{equation}
\sech^2x\sim-\frac{1}{(x-x_*)^2}.
\label{eq:sech-pole}
\end{equation}
For the integer hierarchy in the form
\begin{equation}
\psi''+\left[\epsilon^{-2}+m(m+1)\sech^2x\right]\psi=0,
\label{eq:soliton-epsilon}
\end{equation}
introduce $y=(x-x_*)/\epsilon$. The local equation becomes
\begin{equation}
\frac{\dd^2\psi}{\dd y^2}
+\left[1-\frac{m(m+1)}{y^2}\right]\psi=0.
\label{eq:cluster-equation}
\end{equation}
Its solutions are
\begin{equation}
\psi(y)=\sqrt{y}\,H_{m+1/2}^{(1,2)}(y).
\label{eq:half-integer-Hankel}
\end{equation}
For half-integer order, these functions reduce exactly to $e^{\pm\ii y}$ multiplied by finite polynomials in $1/y$. Their large-$y$ expansions terminate and have no conventional Airy-type Stokes switching. This is the fake-Airy mechanism identified by Berry and Howls \cite{BerryHowls1990}.

Equation~\eqref{eq:B-exact} is the global completion of that local result. If the zero--pole cluster is replaced by a single isolated turning point, one obtains a spurious nonzero reflected exponential. If the cluster is treated exactly, its off-diagonal connection coefficient is proportional to $\sin(\pi\lambda)$ and vanishes for the Darboux hierarchy. The local termination and the global reflection zero are therefore the same fact expressed in two different bases.

\section{Divergent iteration about an exact fixed point}

The remaining question is why Berry's iterative diagonalization can diverge when an exact one-way solution exists. Write an exact traveling-wave ansatz as
\begin{equation}
\psi(x)=K^{-1/2}(x)
\exp\left[\frac{\ii}{\epsilon}\int^xK(s)\,\dd s\right].
\label{eq:phase-integral-ansatz}
\end{equation}
Substitution into Eq.~\eqref{eq:wave-equation} gives the exact nonlinear relation
\begin{equation}
P^2=K^2-
\epsilon^2K^{1/2}\left(K^{-1/2}\right)''.
\label{eq:exact-K}
\end{equation}
A natural fixed-point iteration is
\begin{equation}
K_{n+1}=\mathcal F[K_n]
\equiv
\left[
P^2+
\epsilon^2K_n^{1/2}\left(K_n^{-1/2}\right)''
\right]^{1/2}.
\label{eq:Berry-iteration}
\end{equation}
For the one-soliton, the exact fixed point is
\begin{equation}
K_*(x)=S'(x)
=\frac{1+\epsilon^2}{1+\epsilon^2\tanh^2x}.
\label{eq:K-star}
\end{equation}
Let
\begin{equation}
K_n=K_*e^{\eta_n},
\qquad |\eta_n|\ll1.
\label{eq:eta-def}
\end{equation}
Using
\begin{equation}
K^{1/2}\left(K^{-1/2}\right)''
=-\frac{1}{2}(\ln K)''
+\frac{1}{4}\left[(\ln K)'\right]^2,
\label{eq:logK-identity}
\end{equation}
we find (Appendix B)
\begin{equation}
\eta_{n+1}=\mathcal L\eta_n+O(\eta_n^2),
\label{eq:linearized-iteration}
\end{equation}
with
\begin{equation}
\mathcal L
=-\frac{\epsilon^2}{4K_*}
\frac{\dd}{\dd x}
\left(\frac{1}{K_*}\frac{\dd}{\dd x}\right).
\label{eq:L-operator}
\end{equation}
In the weighted inner product
\begin{equation}
\langle f,g\rangle_*
=\int_{-\infty}^{\infty}K_*f^*g\,\dd x,
\label{eq:weighted-inner-product}
\end{equation}
its quadratic form is
\begin{equation}
\langle f,\mathcal Lf\rangle_*
=\frac{\epsilon^2}{4}
\int_{-\infty}^{\infty}
\frac{|f'|^2}{K_*}\,\dd x\geq0.
\label{eq:L-positive}
\end{equation}
The operator is unbounded. A wave packet of local wavenumber $q$ supported where $K_*$ varies slowly is amplified approximately by
\begin{equation}
\mu(q)\sim\frac{\epsilon^2q^2}{4K_*^2}.
\label{eq:high-q-growth}
\end{equation}
Thus the exact reflectionless fixed point is not an attracting fixed point of the iteration. Repeated differentiation amplifies progressively finer structure, exactly the mechanism behind the factorial growth found in Ref.~\cite{GjajaBhattacharjee1990}. The existence of an exact fixed point does not imply convergence of the algorithm intended to find it. For generic analytic systems, the same complex singularities control both the late terms and an exponentially small transition, and Berry's interpretation is physically correct. The reflectionless hierarchy is exceptional because the singulant survives while its global off-diagonal multiplier vanishes.

\section{What cancels, and what does not}

It is useful to distinguish three statements.

First, the exact connection coefficient multiplying the opposite traveling branch vanishes:
\begin{equation}
B_m(k)=0.
\label{eq:exact-connection-zero}
\end{equation}
This is basis independent once the incoming and outgoing branches have been fixed. The coefficient-level structure of these divergent constructions, including the location of their Borel singularities and the cancellation of their effect in observables, is analyzed in Appendix C.

Second, the finite Darboux solution and the half-integer Bessel comparison function terminate. This termination is representation dependent but exact.

Third, the Berry iterative and Lie-transform constructions can remain infinite and factorially divergent. This divergence is also representation dependent. It does not contradict Eq.~\eqref{eq:exact-connection-zero}.

One should therefore not infer that every local Stokes matrix assigned to a resolved cluster vanishes separately. Local factorization of the connection problem is not unique. In the cluster basis of Berry and Howls there is no off-diagonal switching. In a basis that treats the simple zeros and the pole separately, nonzero intermediate factors may occur, but their ordered product must reproduce Eq.~\eqref{eq:B-exact}. The invariant statement is that the complete off-diagonal Stokes holonomy vanishes.

The diagonal part does not vanish. It is the transmission phase in Eq.~\eqref{eq:transmission-phase}, or equivalently the compact holonomy left after the similarity frame closes. Reflectionlessness cancels the final squeeze, not the phase.

\section{Consequences for classical adiabatic theory in plasma physics}

The results of this paper are broadly applicable. Below, we discuss their implications for classical adiabatic theory in plasma physics. 

\subsection{Magnetic moment and bounce action}

Kruskal's all-orders theorem guarantees a formal invariant for nearly periodic Hamiltonian motion, including gyromotion and mirror bounce motion \cite{Kruskal1962}. The coefficients of that invariant need not converge. The present example shows why one must not identify this divergence automatically with loss of confinement or orbit-class conversion. A divergent guiding-center or bounce-center series states that no single local near-identity transformation can be improved indefinitely. The physical loss of an invariant is controlled by a global connection coefficient across the complete orbit history. In special geometries that coefficient can vanish even though the formal series diverges.

The analogy should not be overstated: realistic guiding-center dynamics has more degrees of freedom and may contain resonances or separatrix crossings absent from Eq.~\eqref{eq:wave-equation}. The rigorous lesson is narrower and more useful: the late terms determine the available exponentially small scale, but a separate global calculation is required to determine its multiplier.

\subsection{Plasma-wave reflection and mode conversion}

Slowly varying plasma waves are commonly treated by ray tracing and wave-action conservation. At an avoided crossing or conversion layer, the local eigenvectors vary rapidly and the WKB basis develops exponentially weak coupling. Equation~\eqref{eq:B-exact} is an exactly solvable example in which the complex singularities and the scale $e^{-\pi/\epsilon}$ remain present, but the net backward-wave amplitude vanishes because the global multiplier is zero. Thus a numerical or asymptotic diagnostic based solely on late-term growth can falsely predict a surviving converted wave.

\subsection{Soliton connection}

The P\"oschl--Teller potentials used here are not arbitrary solvable wells. At integer coupling they are the reflectionless spectral potentials associated with pure KdV solitons. This closes a historical circle. The KdV soliton was identified in the study of nonlinear waves in a collisionless plasma \cite{ZabuskyKruskal1965}, while the all-orders adiabatic framework was developed to understand slowly varying particle motion in plasmas \cite{Kruskal1962}. In the present problem the two themes meet: Darboux soliton structure enforces exact cancellation of the same mode conversion that an adiabatic expansion tries to calculate.

The finite Darboux chain is the globally adapted representation. The divergent Berry or GB Lie-transform sequence is a local representation built from repeated differentiation. That they are not equivalent is a concrete realization of Kruskal's principle that an asymptotic construction must be interpreted through the exact problem it approximates, not mistaken for the exact physics itself.

\section{Discussion and Conclusion}

The result resolves a longstanding interpretive issue in a form that is natural for plasma physics. The 1990 analysis of Gjaja and Bhattacharjee established that Berry iteration and Lie-transform perturbation theory can diverge factorially and argued that the divergence is an artifact of the asymptotic construction rather than direct evidence of transitions \cite{GjajaBhattacharjee1990}. Berry and Howls then showed that reflectionlessness is encoded locally by a tightly organized transition-point cluster whose exact comparison functions possess no ordinary Airy Stokes phenomenon \cite{BerryHowls1990}. The later reflectionless examples demonstrated that the original adiabatic constructions still need not terminate or converge \cite{GjajaBhattacharjee1992}.

The present synthesis identifies the exact object common to all three approaches. Darboux dressing constructs the one-way solution algebraically. Its modulus and phase define an Ermakov similarity frame and an exact action. The frame closes asymptotically, proving zero net squeeze and hence zero conversion. The full hypergeometric connection coefficient vanishes through the factor $\sin(\pi\lambda)$. Yet the same exact frame is an unstable fixed point of derivative iteration. The divergent sequence is therefore not the transition amplitude; it is a failed local representation of a globally regular and transitionless transformation.

This conclusion is consistent with, rather than opposed to, the generic significance of superadiabatic divergence. In most analytic systems, optimal truncation locates the exponentially small scale of a real transition, and exact WKB supplies a nonzero Stokes multiplier \cite{Berry1990Histories,Nenciu1993,Suzuki2024}. What fails is the converse assertion. Divergence identifies where beyond-all-orders information becomes necessary. The global connection data determine whether that information appears as reflection, mode conversion, an endpoint change of action, a phase shift, or an exact cancellation.

The plasma-physics moral is particularly close to Kruskal's viewpoint. Formal adiabatic invariants are local recursive objects. Their divergence need not mean that the underlying orbit has lost its organizing invariant. Conversely, a globally small transition cannot be inferred from low-order apparent convergence. One must distinguish the asymptotic frame from the monodromy of the exact dynamics.

A coefficient-level resurgent completion remains possible. One may construct the Borel transform of the specific canonical generating functions in Ref.~\cite{GjajaBhattacharjee1990} and show explicitly that their lateral discontinuities reconstruct a diagonal connection matrix with zero off-diagonal entry. Borel summation of formal adiabatic invariants is known under suitable analyticity assumptions \cite{CostinDupaigneKruskal2004}. Such a reconstruction would expose the cancellation inside the original coefficient families, but it is not required for the observable theorem proved here: the exact connection coefficient, exact similarity frame, and instability of the iteration are already determined independently.

\begin{acknowledgments}
This paper was shaped by correspondence and discussions with Professor Michael Berry. The author acknowledges gratefully his collaboration with his former doctoral student, Dr. Ivan Gjaja, at Columbia University. This work is supported by the Department of Energy and the National Science Foundation.
\end{acknowledgments}

\appendix

\section{Hypergeometric connection formula}

For Eq.~\eqref{eq:general-PT}, $z=(1-\tanh x)/2$ gives
\begin{equation}
\begin{aligned}
f_\lambda^+&=e^{\ii kx}{}_2F_1(a,b;c;z),\\
a&=-\lambda,\qquad b=\lambda+1,\qquad c=1-\ii k.
\end{aligned}
\end{equation}
Since $a+b=1$, the connection formula about $z=1$ reads
\begin{equation}
\begin{aligned}
{}_2F_1(a,b;c;z)
={}&C_1{}_2F_1(a,b;1+\ii k;1-z)\\
&+C_2(1-z)^{-\ii k}\\
&\quad\times{}_2F_1(c-a,c-b;1-\ii k;1-z).
\end{aligned}
\end{equation}
where
\begin{equation}
C_1=\frac{\Gamma(c)\Gamma(-\ii k)}
{\Gamma(c-a)\Gamma(c-b)},
\qquad
C_2=\frac{\Gamma(c)\Gamma(\ii k)}
{\Gamma(a)\Gamma(b)}.
\end{equation}
As $x\rightarrow-\infty$, $1-z\sim e^{2x}$, so the two terms become $e^{\ii kx}$ and $e^{-\ii kx}$, respectively. This gives Eqs.~\eqref{eq:A-lambda} and \eqref{eq:B-gamma}. The reflection identities
\begin{equation}
\Gamma(-\lambda)\Gamma(1+\lambda)
=-\frac{\pi}{\sin(\pi\lambda)},
\end{equation}
\begin{equation}
\Gamma(1-\ii k)\Gamma(\ii k)
=-\frac{\ii\pi}{\sinh(\pi k)}
\end{equation}
then give Eq.~\eqref{eq:B-exact}.

\section{Linearization of the phase-integral iteration}

Let
\begin{equation}
G[K]=K^{1/2}\left(K^{-1/2}\right)''
=-\frac12(\ln K)''+\frac14[(\ln K)']^2.
\end{equation}
For $K=K_*e^\eta$,
\begin{equation}
\delta G=-\frac12\eta''+\frac12(\ln K_*)'\eta'.
\end{equation}
Since
\begin{equation}
K_{n+1}^2=P^2+\epsilon^2G[K_n],
\end{equation}
and $K_*^2=P^2+\epsilon^2G[K_*]$, one obtains
\begin{equation}
\begin{aligned}
2K_*^2\eta_{n+1}
=\epsilon^2\bigg[&-\frac12\eta_n''\\
&+\frac12(\ln K_*)'\eta_n'\bigg],
\end{aligned}
\end{equation}
which is Eq.~\eqref{eq:L-operator}.

\section{Borel structure of the divergent constructions}
\label{app:gb-borel}
\label{fig:label_verification}

The results of Secs.~IV--IX determine the connection problem independently 
of any resummation. This appendix asks where,
inside the divergent representations themselves, the cancellation
resides. Four statements answer the question: (a)~the exact phase-integral 
function is analytic in $\epsilon$, so nothing about it \emph{requires} resummation; (b)~its
formal $\epsilon$-expansion converges precisely at integer coupling,
and at non-integer coupling diverges with the exact reflection
amplitude as the coefficient of the factorial growth; (c)~the
Lie-transform families, by contrast, carry genuine Borel
singularities at the reflection scale even at integer coupling; and
(d)~the effect of those singularities nevertheless cancels
identically in the connection matrix.

\emph{(a)The exact object is analytic.} From Eq.~(51),
$K_{*}(x;\epsilon)=(1+\epsilon^{2})/(1+\epsilon^{2}\tanh^{2}x)$ is
singular only at $\epsilon=\pm i/\tanh x$, and is therefore analytic
in the disk $|\epsilon|<1$ uniformly on the real line. The radius is
fixed by the bound state at $k=i$---the poles of the transmission
amplitude~(18)---not by the turning-point cluster of Sec.~VII.
Whatever diverges in the sequel is thus a property of a
representation, not of the object represented.

\begin{figure*}[t]
\centering
\includegraphics[width=\textwidth]{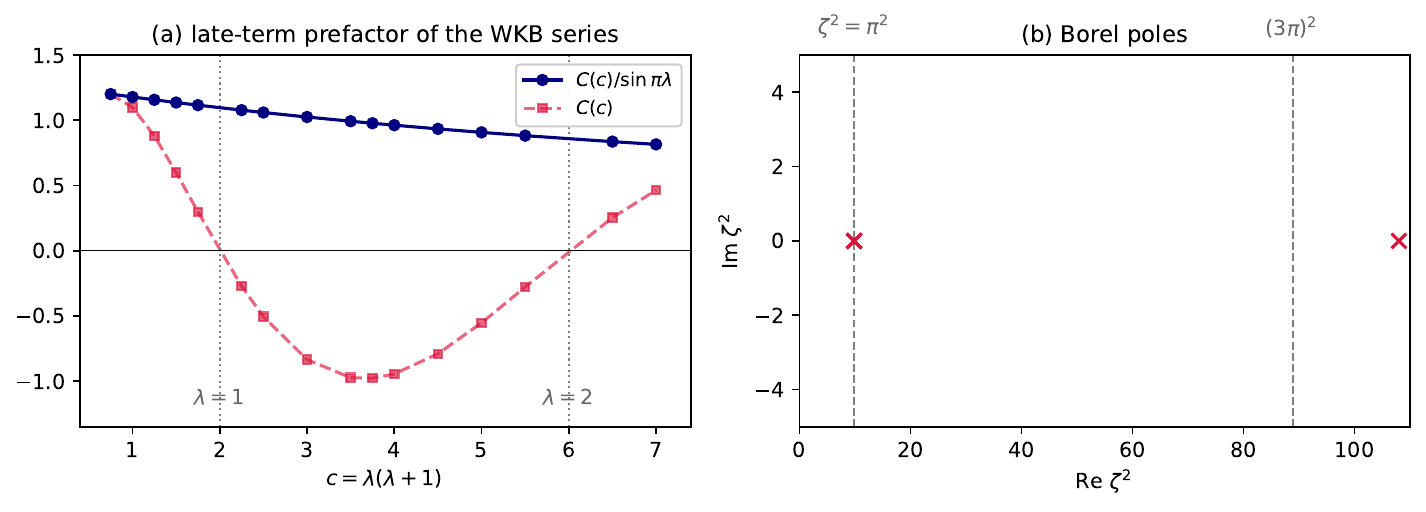}
\caption{Borel structure of the divergent constructions. (a) Late-term prefactor $C(c)$ of the WKB expansion: the ratio $C(c)/\sin(\pi\lambda)$ is smooth and nonzero through the reflectionless couplings $c=2$ and $c=6$. (b) Borel poles of the $n=3$ Lie-transform family in the $\zeta^2$ plane: three approximant poles coalesce near $\zeta^2=\pi^2$, with the next singularity near $\zeta^2=(3\pi)^2$.}
\label{fig:borel_verification}
\end{figure*}

\emph{(b)Late terms of the phase-integral expansion}. Solving the exact relation~(49) recursively for
$P^{2}=1+c\,\epsilon^{2}\,\mathrm{sech}^{2}x$ with
$c=\lambda(\lambda+1)$ gives the formal expansion
$K=1+\sum_{n\ge1}\kappa_{n}(x)\,\epsilon^{2n}$. At integer coupling
this series \emph{converges}: for $m=1$ it terminates at $x=0$,
where $K(0)=1+\epsilon^{2}$, and for $m=2$ it sums to a rational
function, with radius of convergence set by the bound-state poles as
above. At noninteger coupling the same recursion yields factorial
growth,
\begin{equation}
\kappa_{n}(0)\;\sim\;C(c)\,\frac{\Gamma(2n+b)}{\pi^{2n}},
\qquad
C(c)\;=\;g(c)\,\sin(\pi\lambda),
\label{eq:lateterms}
\end{equation}
with $g$ smooth and nonvanishing. The nearest Borel singularity
therefore lies at $\zeta=\pi$, and the prefactor has the same $\sin(\pi\lambda)$ zero as 
the exact global connection coefficient in Eq. (26), whose large-$k$ form is Eq. (31). We have verified
Eq.~\eqref{eq:lateterms} numerically through order $\epsilon^{32}$
and through both zeros $\lambda=1,2$
[Fig.~\ref{fig:borel_verification}(a)]. For this representation, then, the
cancellation is stronger than resummability: the series simply
converges wherever the potential is reflectionless.
\emph{(c)Borel singularities of the Lie-transform families.}
The generating-function representation is genuinely obstructed. The
large-order law of Ref.~\cite{GjajaBhattacharjee1992} [Eqs.~(10) and
(11) there] fixes the late terms of the $n=3$ family completely: the
$2^{-k}$ prefactor of their Eq.~(10) cancels the $2^{k+1}$ of their
Eq.~(11), and the phase factors collapse to a fixed sign, giving,
for $k$ odd,
\begin{equation}
w_{3k}\;\simeq\;12\,I\sin2\theta\;
\frac{(k+2)!}{\pi^{k+3}}\,\lambda_{D}(k+3),
\label{eq:w3k}
\end{equation}
where $\lambda_{D}(s)=\sum_{j\ge0}(2j+1)^{-s}$. The Borel transform
of the family follows in closed form at leading order,
\begin{equation}
\mathcal{B}_{3}(\zeta)\equiv\sum_{k\ \mathrm{odd}}
\frac{w_{3k}}{k!}\,\zeta^{k}
\;\simeq\;12\,I\sin2\theta
\left[(\pi-\zeta)^{-3}-(\pi+\zeta)^{-3}\right],
\label{eq:borel3}
\end{equation}
a third-order pole at $\zeta=\pm\pi$---precisely the reflection
scale of Eq.~(31)---with a tower at odd multiples of $\pi$ supplied
by $\lambda_{D}$. We have verified Eq.~(11) of
Ref.~\cite{GjajaBhattacharjee1992} exactly through $k=25$, and
confirmed the triple pole at $\zeta^{2}=\pi^{2}$ by Borel--Pad\'e
approximation to six digits [Fig.~\ref{fig:borel_verification}(b)]. The lateral
ambiguity of this family alone, of order
$I\sin2\theta\,e^{-\pi/\epsilon}$, is squeeze-type and
nonzero---even though the exact reflection vanishes.

\emph{(d)Why the generator ambiguities cannot survive in the connection matrix.} At integer coupling, 
the lateral Borel ambiguities of the Lie-transform families---nonzero by Eq.~\eqref{eq:borel3}---cancel 
identically in the connection matrix. The following argument shows why such ambiguities cannot appear in the exact frame-invariant connection data, although it does not resolve their family-by-family resurgent cancellation. The exact connection data 
at integer coupling are analytic at $\epsilon=0$: the off-diagonal coefficient vanishes identically,
$\beta(\epsilon)\equiv0$, and the diagonal datum is the phase
$\delta_{m}(\epsilon)=2\sum_{n=1}^{m}\arctan(n\epsilon)$, analytic
with radius $1/m$. The Hamiltonian is analytic and bounded in a
strip about the real axis (Sec.~II), so the exponential estimates
for analytic averaging \cite{Neishtadt1984,
CostinDupaigneKruskal2004} apply: the optimally truncated Lie
construction approximates the exact evolution to
$O(e^{-c/\epsilon})$, and the formal power series it assigns to any
frame-invariant quantity is Gevrey-1 asymptotic to the exact
quantity. An asymptotic power series of a function analytic at the
origin is unique and coincides with its Taylor series; hence the
observable series converge---the $\beta$-series vanishes term by
term, and the $\delta$-series is the convergent arctangent series.
The factorial divergence, and with it the Borel singularities at
$\zeta=(2j+1)\pi$, are therefore confined to the generators of the
frame transformation: they lie in the kernel of the composition
producing the connection matrix, where their lateral ambiguities
cancel identically.

\end{document}